\documentstyle[12pt]{article}
\begin{document}

\newcommand{\nd}[1]{/\hspace{-0.5em} #1}
\begin{titlepage} 
\bigskip
\hskip 3.7in\vbox{\baselineskip12pt
\hbox{SWAT-182}
\hbox{hep-th/yymmddd}}
\bigskip\bigskip\bigskip\bigskip

\centerline{\large \bf GENERAL ABELIAN ORIENTIFOLD MODELS}
\centerline{\large \bf AND ONE LOOP AMPLITUDES}
\bigskip\bigskip
\bigskip\bigskip

\centerline{D\'{o}nal O'Driscoll} 
\bigskip\bigskip

\centerline{Department of Physics, University of Wales Swansea }
\centerline{Singleton Park, Swansea, SA2 8PP, UK}
\centerline{{\em pydan@swansea.ac.uk}}
\bigskip\bigskip

\begin{abstract}
\baselineskip=16pt
We construct a one loop amplitude for any Abelian orientifold point group for 
arbitary complex dimensions. From this we show several results for orientifolds 
in this general class of models as well as for low dimensional 
compactifications. We also discuss the importance and structure of the 
contribution of orientifold planes to the dynamics of D-branes, and give a 
physical explaination for the inconsistency of certain $Z_4$ models as 
discovered by Zwart.
\end{abstract}
\end{titlepage}

\section{Introduction}
Though there has been much excitement about $D$--branes in the past few years, 
it has only been more recently that 
the importance of orientifolds \cite{dlp} in these models has been appreciated 
\cite{djm} \cite{s}. An orientifold can stand for two related things, a 
compactification model or a plane. A natural way of looking at the latter is as 
the unorientated counterparts of the $D$--branes. As they are unoriented they 
have somewhat different properties and are undynamical, but they can not be 
discounted as they can couple to $D$--branes via open and unoriented closed 
strings. Indeed the open type I string can be considered as a twisted sector of 
the closed string under these conditions. 

Orientifold models arise when we combine the worldsheet parity symmetry with an 
orbifold point group. The fixed points in the target space then become the fixed 
orientifold plane. These planes carry RR charges which in turn require the 
presence of $D$--branes to cancel. The $D$--branes and $o$--planes interact via 
open strings which have Chan--Paton factors associated with them, and hence also 
Wilson lines. In the T--dual models which produce the $D$--branes from string 
theory the Wilson lines describe the position of the $D$--branes in the compact 
space, and thus play a significant roll in their dynamics by giving the gauge 
groups associated with the $D$--branes. When this is combined with an discrete 
modding to produce a orbifold space the $D$--branes become identified under the 
point group so that not all of the are dynamical. This is reflected in the 
Wilson lines. In order to understand how the Wilson lines behave under the 
action of the point group we require the action of the point group on the 
Chan--Paton factors. This we do by investigating the open string one loop 
amplitudes in the T--dual model.

The aim of this paper is to find a general solution to the action of the point 
group on the Chan--Paton factors for the set of Abelian point groups defined on 
complex compact coordinates. Though these are not the most general orbifold 
models, they do form the most important class. Much work has been performed in 
this area already and this paper hopes to finish part of that project. Some 
general results, independant of the number of dimensions being compacted, are 
produced and we discuss some of the problems that have arisen in the consistancy 
of these models. We also show how strongly the dynamics of the $D$--branes are 
in fact controlled by the presence of the orientifold planes. They do not simply 
increase the number of possible gauge groups that $D$--branes can have, but 
tightly control the way they can move together by defining with unexpected 
rigidity the action of the point group on the Chan--Paton factors and Wilson 
lines.

In section 2 we outline the theory and features we wish to employ as well as our 
notation. Some of the notation will change for different parts of the theory but 
this is to emphaise certain patterns in the amplitudes. In section 3 we 
calculate the one loop open string amplitudes we will need \cite{gs, cp}, giving 
them in the 
most general form possible; while in section 4 we extract the divergences from 
them. These divergences, in fact tadpole equations, are the charge 
cancellation constraints in the dual model and are what allow us to solve for 
the action of the point group on the Chan--Paton factors. Surprisingly enough, 
we are able to find a very general formula for this, which is what gives us the 
ability to discuss the strong influence of the orientifold planes. Also in 
section 4 we take a look at orientifold models which have been compacted down to 
two dimensions. In section 5 we use some of the results of section 4 to give a 
physical explaination for the strange inconsistancy of certain $Z_4$ point group 
models as recently discovered by Zwart. Finally we take a quick look at how the 
orientifold groups can be extended.

\section{The Amplitudes I: Features}
In this section we start to write down an scheme for constructing the Klein 
bottle, M\"{o}bius strip and cylinder amplitudes for the open string. We follow 
the conventions of Gimon and Polchinski \cite{gp}, and use the work of Gimon and 
Johnson \cite{gj} as our basis. Most of the background work and theory can be 
found in refs \cite{gp, gj, dp1, dp2, bl, z, fg, abpss} and we will leave the 
reader to consult those papers for the relevant material. 

Here we shall just give the basic tools and definitions of what we are 
calculating. Our ultimate aim is to produce a set of equations that will allow 
us to write down the conditions on the projective representation of the 
orientifold point group acting on the Chan Paton factors for any given model.

\subsection{General Formulae}
The complete amplitude we wish to calculate can be written as
\begin{equation} 
\int^\infty_0{dt\over t} \left[Tr_c({\bf P}(-1)^{\bf F}e ^{-2 \pi t(L_0 + 
\bar{L_0})} +  Tr_o({\bf 
P}(-1)^{\bf F}e ^{-2 \pi tL_0})\right]
\end{equation}
We use {\bf P} here to denote the generalized GSO projector extended to include 
the point group of the orientifold model and {\bf F} is the space time fermion 
number. t is the loop modulus for the cylinder; it can be related to the 
cylinder length, to give a common length scale, by $t= 1/2l, 1/8l, 1/4l$ for the 
cylinder, M\"{o}bius strip and Klein bottle diagrams respectively. $L_0$ 
represents the hamiltonian of the string in the various twisted and untwisted 
sectors. 

\subsection{General Point Groups}
At this point we are going to relax the constraints of supersymmetry and modular 
invariance to allow a larger range of abelian point group models. Since we 
desire a degree of arbitariness for the number of compact dimensions in our 
analysis this 
extends in to the lattices used for compactification  and hence the allowed 
automorphisms. Though for the most part the actual group structure of the 
lattice is not as important as the point group we have increasing number of 
possiblities as the number of compact dimensions increase. However, for this 
paper we will restrict our discussion to the purely Abelian 
case on complex coordinates as these can be related in many cases to Calabi-Yau 
spaces. We will denote the 
$k$th element of the Abelian groups as   $\alpha^k_N$ where $\alpha_N$ is the 
generator. Thus the elements of the 
orientifold group can be written as
\begin{equation} \{{\bf 1}, \Omega, \alpha^k_N, \Omega\cdot\alpha^k_N \} 
\end{equation}
The extension to point groups with more than one generator is obvious.

Without loss of generality we can  divide our compact space up using complex 
coordinates, $z_i$.  The reason for this is that the $T^2$ has two possible Lie 
algebra valued lattices corresponding to the root diagrams for $SU(3)$ and 
$SU(2) \times SU(2)$ and allows the incorporation of all Abelian point groups of 
interest. We divide our spacetime into three sections:
\begin{itemize}
\item The compact coordinates $I = 1, ..., 2d$.
\item The uncompact coordinates $\mu = 1, ..., 2\nu$.
\item The ghost coordinates which swallow two dimensions (corresponding to the 
light cone quantization).
\end{itemize}
If the total number of dimensions in the theory is D then we must have
\begin{equation} 
2\nu + 2d + 2 = D 
\end{equation}
where D in the case of superstring theory is 10. 

Because we are complexifying our space we can define the action of each element 
of our point group on a torus 
independently of the rest to find the amplitudes, though the final overall 
results will take into account all 
sectors and the overlap between the action of the point groups. 

Take a general cyclic group $Z_N$ with element $\alpha^k_N$; following the 
literature we define its action on 
a particular complex coordinate, say $\rho_1 =x^8+ix^9$ as:
\begin{equation}
\alpha^k_N: \rho_1 \rightarrow e^{\pm{2\pi in\over N}}\rho_1
\end{equation}
for the bosonic and NS sectors; while for the R sector the point groups acts on 
the appropriate fermions with the action:
\begin{equation}
 \alpha^k_N: e^{\pm{2\pi ik\over N}J_{89}} 
\end{equation}

Note that when the point group is of the form $Z_N \times Z_M$, which we will 
call a {\em cross} group as opposed to a {\em pure} group otherwise, we can 
define the action of elements of both groups, generated by $\alpha^n_N$ 
and $\alpha^m_M$ on a single complex coordinate in a combined form, i.e. 
\begin{equation}  
\rho_1 \rightarrow e^{2\pi in\over N}e^{\pm{ 2\pi im\over M}}\rho_1 
\end{equation}
In all cases we must choose $\pm$ a priori. Models with point groups of the form 
$Z_N \times Z_M \times Z_P$ act on a single complex coordinate are not allowed 
to occur, though more complex actions where each group acts on a seperate 
but overlapping pair of coordinates are allowed.

\subsection{Dp-Branes}
Since we are dealing with T-dual models we have the presence of $D$-branes in 
the theory corresponding to open strings with Dirichlet boundary conditions. 
These make the open string one loop amplitudes much more interesting but 
unfortunately reduce the generality of our expressions. In Type I string theory, 
taken as a orientifold of type IIB theory the eigenvalues of the parity operator 
$\Omega$ satisfy \cite{gp}
\begin{equation}  
\Omega^2 = (\pm i)^{(9-p)/2} 
\end{equation}
This has values of $\pm1$ for $p=1, 5, 9$ which is consistant with what is known 
about Type I theory. Other values of $p$ get projected out. As a result of this 
we can not use arbitary $D$ above but must set it equal to 10 for superstrings. 
It is not obvious to us how to generalize this expression with this technique 
but as there is no need to we shall leave it as it stands. Later we shall see 
that the amplitudes do seem to see this value for the total dimension.

A side consequence of this form for the $\Omega^2$ is that the group nature of 
the projection representation of the point group and thus the Chan Paton factors 
associated with each $D$-brane via the open string interactions is fixed. The 
$D$9-brane and the $D$1-brane have orthogonal factors, $SO(n)$, while the 
$D$5-brane symplectic factors, $Sp(n)$. See \cite{gp, tasi, w} for further 
discussion on this point.

The $D9$-Brane is always present as it corresponds to the action of the 
worldsheet parity, $\Omega$ with the identity of the point group. It is 
essentially the open string with all Neumann boundary conditions and is required 
to cancel the presence of the $o9$--plane induced in the theory by $\Omega$. 
Other members of the point group determine which other $o$-planes can exist. 
These latter objects carry RR charge which needs to be cancelled by the opposite 
charge appearing on the $Dp$-branes. However, depending on the action of the 
point 
group and hence the conserved charge not all these will be present in the 
theory\footnote{we are grateful to Stefan F\"{o}rste for help with this point}. 
For example, the $D1$-brane only appears in low dimensional compactification, 
while $D5$-branes are not present in $T^8/Z_2$ but are present in $T^8/Z^n_2$ 
for $n=2, 3$ \cite{fg}.

A point to note is that for five branes there are different configurations 
possible under the various point groups so that there are different types giving 
rise to sectors of interaction between $5_i5_j, 95_i, 5_i9, 15_i, 5_i1$ 
according to the model. This is a feature of the way in which we choose the 
models to act on the complex coordinates.  

\subsection{Compact Zero Modes}
In dealing with a variation of standard orbifold models there are twisted and 
untwisted sectors in the theories. In the former the bosonic zero modes, 
corresponding to discrete momentum and winding factors vanish, but they must be 
accounted for in the untwisted sector. There are various types of contribution 
to consider depending on the $D$-branes involved. Following Berkooz and 
Leigh\cite{bl} we will write their 
contributions in the form 
\begin{eqnarray}
M_j= (\sum_{n \in {\bf Z}}e^{-\pi t n^2 /v_j})^2 \\
W_j= (\sum_{\omega \in {\bf Z}}e^{-\pi t \omega^2 v_j})^2 
\end{eqnarray}   
where the momentum sum is on the root lattice ${\bf \Gamma}$; and the winding 
modes are summed over the dual 
lattice ${\bf \Gamma}\star$. $v_j$ is the volume of the particular torus.  
   
\subsection{Chan Paton Factors}
Chan Paton factor are intrinsically features of the open string, and as such 
will not appear in the Klein 
bottle expressions. Traditionally they have provided gauge groups for the open 
string but in the dual models 
they play a very fundamental role and are responsible for a lot of the 
interesting features of the theory. 

The action of the orientifold group on the Chan Paton factors, $\lambda$, should 
form a projective 
representation of the group up to a phase controlled by the various $D$-brane 
sectors they are being taken 
in. If {\bf P} represents an element of the point group and $p$ the $D$-brane 
sector, then the projective 
representation of the orientifold group is given by the matrices $\gamma$ 
satisfying:
\begin{eqnarray}
\lambda \propto \gamma_{{\bf P}, p}\lambda\gamma^{-1}_{{\bf P}, p} \\
\lambda \propto \gamma_{\Omega, p}\lambda^T\gamma^{-1}_{\Omega, p} 
\end{eqnarray}
What appears in the one loop amplitudes for the M\"{o}bius strip and cylinder 
are the trace over the 
different $\gamma$'s according to sector. Thus by demanding cancellation of 
tadpoles there are constraints on the $\gamma$'s which in turn will allow 
specific representations to be formed which are unique upto unitary 
transformations on the Chan--Paton factors \cite{bl}. 

\section{Amplitudes II: The Equations}

\subsection{General Features of the Solutions}
The amplitudes we are specifically calculating are 
 \begin{eqnarray} 
{\rm KB}:Tr^{U+T}_{NSNS+RR}{ {\Omega\over2}{\bf P}{1+(-1)^{\bf F} \over 2}
e^{-2\pi t(L_0+{\bar L_0})}}    \\
{\rm MS}:Tr_{NS-R}^{\lambda\lambda}{{\Omega\over2}{\bf P}{1+(-1)^{\bf F} \over 
2} 
e^{-2\pi tL_0}} \\ 
{\rm C}:Tr_{NS-R}^{\lambda\lambda^\prime}{{1\over2}{\bf P}{1+(-1)^{\bf F} \over2 
}
e^{-2\pi tL_0}} ,
\end{eqnarray}
where the point group projection operator {\bf P} is given by 
\begin{equation} 
{1 \over N} \sum_{k=0}^{N_i -1} \alpha^k_{N_i} 
\end{equation}
N is the order of {\bf P} and $N_i$ is the order of the subgroups where 
applicable.

In each amplitude there is a sum over three sets of theta functions. In order 
these will represent the contributions from the Neveu--Schwarz and Ramond 
sectors. The theta functions 
appearing with negative index are the bosonic contributions with the factors of 
$f(t)$ (and in 
some of the expressions 
trignometric contributions depending on $z_i$) being present to allow the 
rewriting in terms of the theta 
functions. There also appears in each amplitude a factor proportional to 
$t^{-({D-2d \over 2})}$ which is the 
uncompactified momentum contribution. This is rewritten in simpler form as 
$t^{-(\nu +1)}$ or equivalently 
$t^{(d-5)}$ using $D=10$.

The types of $D$-brane the open string ends on is given by the labels $\lambda$ 
and $\lambda^\prime$. For the 
M\"{o}bius strip we have $\lambda=\lambda^\prime$, that is $99, 5_i 5_i, 11$ 
sectors. For the cylinder we 
must take into account all possible combinations of $D$-branes that can appear 
in the particular model under 
construction eg
\[ 99,\, 95_i,\, 91,\, 5_i5_j (+ 5_i5_i),\, 5_i1,\, 11 \]
The Klein bottle ends only on $o$-planes so does not receive these contributions

Though so far we have been very general in our approach we have yet to 
incorporate the Gimon--Polchinski consistancy conditions. They affect the 
presence of the Klein bottle and M\"{o}bius strip in the different twisted 
sectors.
 
\subsection{Conventions}
We will express the amplitudes using the Jacobi $\vartheta$--functions for the 
twisted sectors and the $f_i$ functions for the untwisted sectors as this allows 
us to easily see the relevant patterns for a general point group. For the open 
string amplitudes, the $D$9-branes correspond to the string with Neumann 
boundary conditions on both ends. These will give a contribution to the traces 
according to
\begin{equation} 
Tr(e^{\pm {2 \pi i k \over N_i}}) = (4 sin^2 {\pi k \over N_i})^{-d^\prime} 
\end{equation}
where $d^\prime$ means we take into account every complex dimension the group 
$Z_{N_i}$ is acting apon. 

For our theta functions we use the definitions as 
\begin{eqnarray}
\vartheta_1(z|t) & =&2q^{1/4} \sin \pi z \prod_{n=1}^{\infty} 
(1-q^{2n})\prod_{n=1}^{\infty} 
(1-q^{2n}e^{2\pi iz})\prod_{n=1}^{\infty} (1-q^{2n}e^{-2\pi iz})\nonumber  \\
\vartheta_2(z|t) & =&2 q^{1/4} \cos \pi 
z\prod_{n=1}^{\infty}(1-q^{2n})\prod_{n=1}^{\infty} (1+q^{2n}e^{2\pi 
iz})\prod_{n=1}^{\infty} (1+q^{2n}e^{-2\pi iz}\nonumber) \\
\vartheta_3(z|t) & =& \prod_{n=1}^{\infty} (1-q^{2n})\prod_{n=1}^{\infty} 
(1+q^{2n-1}e^{2\pi iz})\prod_{n=1}^{\infty} (1+q^{2n-1}e^{-2\pi iz})\nonumber \\
\vartheta_4(z|t) & =&\prod_{n=1}^{\infty} (1-q^{2n})\prod_{n=1}^{\infty} 
(1-q^{2n-1}e^{2\pi iz})\prod_{n=1}^{\infty} (1-q^{2n-1}e^{-2\pi iz}) 
\end{eqnarray}
with  $q=e^{-\pi t}$ and $z_i = {k \over N_i}$. 

Also required is
\begin{eqnarray}
f_{1}(q) = q^{1/12} \prod_{n=1}^\infty \left(1-q^{2n}\right),\qquad
f_{2}(q) = q^{1/12} \sqrt{2}\,\prod_{n=1}^\infty \left(1+q^{2n}\right)
\nonumber\\ 
f_{3}(q) = q^{-1/24} \prod_{n=1}^\infty \left(1+q^{2n-1}\right),\qquad
f_{4}(q) = q^{-1/24}\prod_{n=1}^\infty \left(1-q^{2n-1}\right) .
\end{eqnarray}

\subsection{Klein Bottle}
{\em (a)Untwisted sector}
\begin{eqnarray}
\int_0^\infty {dt\over t}(4\pi^2\alpha^\prime t)^{d-5}
{f^8_4(2t)\over f^8_1(2t)} \times {\cal F}(M_i, W_j)
\end{eqnarray}
Here ${\cal F}(M_i, W_j)$ stands for the zero mode contribution which is
\begin{itemize}
\item $\prod_{j=1}^{d} M_j+ \prod^d_{j}W_j$ if the point group consists of a 
single $Z_N$ acting on all the compact coordinates simultaneously.
\item $\prod_{j=1}^{d} M_j + \sum_{\bf 5} \prod^d_i M_i\prod^d_{j \neq i}W_j $ 
when there are 
$D9$- and $D5$-branes in a cross theory. 
\item $\prod_{j=1}^{d} M_j + \sum_{\bf 5} \prod^d_i M_i\prod^d_{j \neq i}W_j + 
\prod^d_{j}W_j$ when all possible branes, including $D$1-branes are present in 
the theory. 
\end{itemize}

We use $\sum_{\bf 5}$ to denote the fact that we must sum over the possible ways 
we can include $D$5-branes in the theory which will be dependant on the action 
of the point group on the compact coordinates\footnote{A quick rule of thumb, 
though not entirely correct, is that $\prod_{j=1}^{d} M_j$ is from the 
$D$9-branes, $\prod^d_i M_i\prod^d_{j \neq i}W_j$ is from the presence of 
$D$5-branes, while $\prod^d_{j}W_j$ is normally present if there are $D$1-branes 
in the theory.}.\\
{\em (b)Twisted sector}\\
In the twisted sectors we need to define the action on the modulus of the torus 
$t_i$. We use the notation 
\begin{equation} t_i = t + \zeta t, \,\,\,  \zeta=\zeta(i) \end{equation}
where the twist $\zeta{=}(m-n)/N_i$ in the closed string channel. For cross 
models we have the same only now the denominator is the larger of the Abelian 
point groups where applicable and the numerator contributions are appropriately 
normalized. That is if the relevant point group is $Z_N \times Z_M$ with $M=aN$ 
such that $a \in {\bf Z}^+$, then 
\begin{equation} \zeta = \pm \left( \frac{m}{M} - \frac{n}{N}\right) = \pm 
\frac{m - an}{M} \end{equation}
In the $T^4/Z_N$ model of Gimon and Johnson \cite{gj} then this simplifies to 
setting $t_i$ to $t^+=t+\zeta 
t,\, t^-=t-\zeta t$ appropriately for the seperate tori. The amplitude is:
\begin{eqnarray} 
\sum_{n=1}^{N-1} 
\int^\infty_0{dt\over t}(4\pi^2\alpha^\prime t)^{d-5}\,\,f_1^{-3\nu}(2t) \times 
\nonumber\\
\left\{-\vartheta^\nu_4(0|2t)\prod_{i=1}^{i=d}\vartheta_4(z_i|2t_i)
\vartheta^{-1}_1(z_i|2t_i) {(2\sin2\pi (z_i -\zeta t))\over(4\sin^2\pi z_i)} 
\right.
\nonumber\\ 
+\vartheta^\nu_3(0|2t)\prod_{i=1}^{i=d}\vartheta_3(z_i|2t_i)
\vartheta^{-1}_1(z_i|2t_i) {(2\sin2\pi (z_i -\zeta t)) \over(4\sin^2\pi z_i)}
\nonumber\\
\left.-\vartheta^\nu_2(0|2t)\prod_{i=1}^{i=d}\vartheta_2(z_i|2t_i)
\vartheta^{-1}_1(z_i|2t_i) {(2\sin2\pi (z_i -\zeta t)) \over(4\sin^2\pi z_i)} 
\right\} 
\end{eqnarray}

\subsection{M\"{o}bius Strip}
{\em (a)Untwisted Sector}
\begin{eqnarray}
 - \int^\infty_0{dt\over t}(8\pi^2\alpha^\prime 
t)^{d-5}\frac{f_{2}^{8}(2t)f_{4}^{8}(2t)}{f_{1}^{8}(2t)f_{3}^{8}(2t)} \times 
{\cal F}(M_i, W_j)
\end{eqnarray}
where the contribution from the zero modes is similiar to the Klein bottle, only 
now we must take into 
account traces over Chan Paton factors in the amplitude. Thus depending on our 
compacting scheme 
${\cal F}(M_i, W_j)$ equals
\begin{itemize}
\item ${\rm Tr}(\gamma_{\Omega,9}^{-1}\gamma_{\Omega,9}^T)\prod_{j=1}^{d} M_j+ 
\sum_{{\bf P}(i)}{\rm Tr}(\gamma_{\Omega{\bf P},B}^{-1}\gamma_{\Omega{\bf 
P},B}^T)\prod^d_{j}W_j$ if the 
point group consists of a single $Z_N$ acting on all the compact coordinates 
simultaneously. $B$ stands for a 
$D$1-brane when $d=4$ and a $D$5-brane otherwise.
\item ${\rm Tr}(\gamma_{\Omega,9}^{-1}\gamma_{\Omega,9}^T)\prod_{j=1}^{d} M_j + 
\sum_{{\bf P}(i)}{\rm Tr} 
(\gamma_{\Omega{\bf P},5_i}^{-1}\gamma_{\Omega{\bf P},5_i}^T)\prod^d_i 
M_i\prod^d_{j \neq i}W_j $ when there 
are $D9$- and $D5$-branes for the cross theory. 
\item ${\rm Tr}(\gamma_{\Omega,9}^{-1}\gamma_{\Omega,9}^T)\prod_{j=1}^{d} M_j + 
\sum_{{\bf P}(i)}{\rm 
Tr}(\gamma_{\Omega{\bf P},5_i}^{-1} \gamma_{\Omega{\bf P},5_i}^T)\prod^d_i 
M_i\prod^d_{j \neq i}W_j  \\
+{\rm Tr} (\gamma_{\Omega,1}^{-1}\gamma_{\Omega,1}^T)\prod^d_{j}W_j$ when $d=4$. 
\end{itemize}
{\em (b)Twisted Sector}\\
The twisted M\"{o}bius strip amplitudes are\footnote{note that for the 
M\"{o}bius strip we have $q=exp(-2\pi 
t)$} :\\ 
For the $D9$--branes:
\begin{eqnarray} 
-\sum_{{\bf P}(i)}
Tr(\gamma^{-1}_{\Omega{\bf P},9}\gamma^T_{\Omega{\bf P},9})
\int^\infty_0{dt\over t}(8\pi^2\alpha^\prime t)^{d-5}f_1^{-3\nu}(iq) \times 
\nonumber \\
\left\{ \vartheta^\nu_3(iq, 0)\prod_{i=1}^{i=d}\vartheta_3(iq, 
z_i)\vartheta_1^{-1}(iq, z_i){(2\sin\pi 
z_i)\over(4\sin^2\pi z_i)} \right. \nonumber \\ 
-\vartheta^\nu_4(iq, 0)\prod_{i=1}^{i=d}\vartheta_4(iq, z_i)\vartheta_1^{-1}(iq, 
z_i){(2\sin\pi 
z_i)\over(4\sin^2\pi z_i)}\nonumber  \\
\left.-\vartheta^\nu_2(iq, 0)\prod_{i=1}^{i=d}\vartheta_2(iq, 
z_i)\vartheta_1^{-1}(iq, 
z_i){(2\sin\pi 
z_i)\over(4\sin^2\pi z_i)}\right\} 
\end{eqnarray}

For the $D5$--branes, we have to be very careful where we place the brane in the 
compact and 
non--compact space time. We shall assume that the $D5$--branes shall fill the 
uncompact spacetime before the 
compact dimensions. In the case that the number of uncompact dimensions is less 
than that of the five brane 
we can organize the permitted $D5$--branes in the theory in the compact 
dimensions according to the action of 
the point group. This naturally splits the mode expansion, and thus the 
amplitudes, into two: those with DD 
boundary conditions and those with NN boundary conditions. We will use a double 
prime to denote those compact dimensions with a $D5$--brane in them and a double 
prime those without, the product being symbolic over the compact dimensions.

\begin{eqnarray} 
-\sum_{{\bf P}(i)}
Tr(\gamma^{-1}_{\Omega{\bf P},5_i}\gamma^T_{\Omega{\bf P},5_i})
\int^\infty_0{dt\over t}(8\pi^2\alpha^\prime t)^{d-5}f_1^{-3\nu}(iq) \times 
\nonumber \\
\left\{\vartheta^\nu_3(iq, 0)\prod^d_i{\vartheta_4^\prime(iq, 
z_i)\vartheta_3^{\prime\prime}(iq, z_i)(2\cos\pi z_i) 
\over \vartheta^{^\prime}_2(iq, z_i)\vartheta^{\prime\prime}_1(iq, 
z_i)}\right. \nonumber  \\ 
-\vartheta^\nu_4(iq, 0)\prod_i^d{\vartheta_3^\prime(iq, 
z_i)\vartheta_4^{\prime\prime}(iq, z_i)(2\cos\pi z_i)  
\over \vartheta^{^\prime}_2(iq, z_i)\vartheta^{\prime\prime}_1(iq, 
z_i)}\nonumber  \\
\left. -\vartheta^\nu_2(iq, 0)\prod_i^d{\vartheta_1^\prime(iq, 
z_i)\vartheta_2^{\prime\prime}(iq, z_i)(2\cos\pi z_i)  
\over \vartheta^{^\prime}_2(iq, z_i)\vartheta^{\prime\prime}_1(iq, z_i)} 
\right\} 
\end{eqnarray}

For the $D1$-branes present when $d=4$, the equation is essentially the same as 
for the $D9$--branes only the change 
of boundary conditions from NN to DD means a shift in the theta functions and a 
loss of the trace contribution that arose out of the NN sector:
\begin{eqnarray} 
-\sum_{{\bf P}(i)}
Tr(\gamma^{-1}_{\Omega{\bf P},1}\gamma^T_{\Omega{\bf P},1})
\int^\infty_0{dt\over t}(8\pi^2\alpha^\prime t)^{d-5} \times\nonumber  \\
\left\{\prod_{i=1}^{i=d}\vartheta_4(iq, z_i)\vartheta_2^{-1}(iq, z_i)(2\cos\pi 
z_i)\right. \nonumber  \\ 
-\prod_{i=1}^{i=d}\vartheta_3(iq, z_i)\vartheta_2^{-1}(iq, z_i)(2\cos\pi 
z_i)\nonumber  \\
\left. -\prod_{i=1}^{i=d}\vartheta_1(iq, z_i)\vartheta_2^{-1}(iq, z_i)(2\cos\pi 
z_i) \right\}
\end{eqnarray}

\subsection{Cylinder}
{\em (a)Untwisted Sector}\\
The contribution from the 99 cylinders are 
\begin{equation}
(Tr(\gamma_{1,9}))^2\int^\infty_0{dt\over t}(8\pi^2\alpha^\prime 
t)^{d-5}{f_4^8(t) \over f_1^8(t)} \prod_j^d M_j   
\end{equation}
From the $5_i5_i$ cylinders:
\begin{eqnarray}
\lefteqn{\int^\infty_0{dt \over t}(8\pi^2\alpha^\prime t)^{d-5}{f_4^8(t) \over 
f_1^8(t)} \times}\nonumber  \\
&&\sum_i M_i\sum_{a,b \in 5_i}(\gamma_{1,5_i})_{aa}(\gamma_{1,5_i})_{bb} 
\prod_{m_{j \neq i}} \sum_\omega e^{-t(2\pi\omega r_j + X_a^{m_j} - 
X_b^{m_j})^2/2\pi\alpha^\prime}
\end{eqnarray} 
That is there is a momentum contribution from the compact dimensions where the 
$D5_i$-branes live and winding contributions otherwise. For the 11 cylinders 
there will only be winding contibutions so we have:
\begin{eqnarray}
\lefteqn{\int^\infty_0{dt \over t}(8\pi^2\alpha^\prime t)^{d-5}{f_4^8(t) \over 
f_1^8(t)} 
\times} \nonumber \\
&&\sum_{a,b \in 1}(\gamma_{1,1})_{aa}(\gamma_{1,1})_{bb} \prod_{m_{j \neq i}} 
\sum_\omega e^{-t(2\pi\omega r + X_a^{m_j} - X_b^{m_j})^2/2\pi\alpha^\prime}
\end{eqnarray}
{\em (b)Twisted Sector}\\
For the twisted cylinder we have to be careful as there are a variety of 
boundary conditions that will give 
different results according the action of ${\bf P}$. The boundary conditions can 
be grouped as NN, DD and ND 
according to the particular branes being uses; while the action of ${\bf P}$ is 
labelled by $z_i$. Note that 
${\bf P}=1$ can also label the uncompact dimensions as well. Thus with the given 
prefactors listed below there is a sum over three sets of products of four 
$\vartheta$ functions. Following the same 
set of conventions as above we have the following theta functions for each 
contribution from the three 
fermion and boson sectors respectively:
\begin{enumerate}
\item NN,   ${\bf P}=1: \vartheta_3(0|t),\,\vartheta_4(0|t),\,\vartheta_2(0|t) 
,\, f^{-1}_1(t)$
\item NN,   ${\bf 
P}=z_i:\vartheta_3(z_i|t),\,\vartheta_4(z_i|t),\,\vartheta_2(z_i|t),\,\vartheta^
{-1}_1(z_i|t)(2 \sin\pi z_i)$
\item ND,   ${\bf P}=1: 
\vartheta_2(0|t),\,\vartheta_1(0|t),\,\vartheta_3(0|t),\,f^{-1}_1(t)$
\item ND,   ${\bf 
P}=z_i:\vartheta_2(z_i|t),\,\vartheta_1(z_i|t),\,\vartheta_3(z_i|t),\,\vartheta^
{-1}_4(z_i|t)$
\item DD,   ${\bf P}=1: 
\vartheta_3(0|t),\,\vartheta_4(0|t),\,\vartheta_2(0|t),\,f^{-1}_1(t)$
\item DD,   ${\bf 
P}=z_i:\vartheta_3(z_i|t),\,\vartheta_4(z_i|t),\,\vartheta_2(z_i|t),\,\vartheta^
{-1}_1(z_i|t)(2 \sin\pi z_i)$
\end{enumerate}
Note that for NN with ${\bf P}=z_i$ we must also account for the contribution of 
$(4\sin^2\pi z_i)$ from the trace over the 
representation of ${\bf P}$ as noted earlier. In the equation below we will have 
this as being implicit. In the ${\bf P} \neq 1$ sectors all the amplitudes will 
contain a factor of $f_1^{-3\nu}(t)$ to correctly normalize the theta 
functions. Naturally in the case $d=4$ we have $\nu=0$ and this contribution 
becomes 1.

A general cylinder amplitude will be of the form 
\begin{eqnarray}
\sum_{{\bf P}(i)}Tr(\gamma_{{\bf P},B})Tr(\gamma_{{\bf P},C})
\int^\infty_0{dt\over t}(8\pi^2\alpha^\prime t)^{-1}f_1^{-3\nu}(t)\times 
\nonumber  \\
\left\{ \prod_{i=1}^4 {\vartheta_{F_i}(z_i|t) \over \vartheta_{B_i}(z_i|t)}
-\prod_{i=1} {^4\vartheta_{F_i}^\prime(z_i|t) \over 
\vartheta_{B_i}^\prime(z_i|t)}
-\prod_{i=1}^4 {\vartheta_{F_i}^{\prime\prime}(z_i|t) \over 
\vartheta_{B_i}^{\prime\prime}(z_i|t)}\right\}
\end{eqnarray}
Where $B$, $C$ can be a 1, $5_i$ or 9 $D$--brane, $z_i$ can be zero and $F_i$ 
labels the relevant boundary conditions according to $B$ and $C$ for the 
fermions, while $B_i$ labels the corresponding bosons for the same conditions. 
The primes 
simply label the fermion sector. When $B$, $C$ or both are $D5$--branes or 
$D1$--branes then there is a sum over the fixed points of the orbifold group as 
well.

\section{Divergences}
Here we will give a general set of rules that can be derived for arbitary 
compactification (up to 
complexification) then concentrate on developing the more general case in $d=4$. 
For convention we adopt 
\begin{equation} v_L = V_L(4\pi^2 \alpha\prime)^{L/2} \end{equation}
to represent the appropriate volume expressions assosiated with each 
compactification scheme.

Taking the limit $t \rightarrow 0$ and using the relevant correspondances 
between $t$ and $l$ we can 
calculate the tadpole divergences from each diagram in each sector. In the 
untwisted sector all dependance on 
$\nu$ and $d$ vanishes in this limit giving us a set of relations for each type 
of brane that holds in all 
cases. For the $D5_i$-branes and $D1$-brane the presence of ${\bf P}(i)$ 
represents the set of 
transformations that defines the branes. In fact the ${\bf P}(i)$ in question 
will be an action of a single 
point group element of order two (if there is no such element then the brane 
will not exist in the model as 
can also be seen by consideration of charge cancellation), so we can use the 
Gimon--Polchinski condition for 
an element $g$ of the point group 
\begin{equation} \gamma^T_{\Omega g}= \pm \gamma_{\Omega g} \end{equation}
These relations, dropping common factors, are: \\
(a) for the $D9$-brane, 
\begin{equation} Tr(\gamma_{0,9})^2 - 
64Tr(\gamma^{-1}_{\Omega,9}\gamma^T_{\Omega,9}) + 32^2 
\end{equation}
(b) for the $D5_i$-branes,
\begin{equation} Tr(\gamma_{0,5_i})^2 - 64Tr(\gamma^{-1}_{\Omega{\bf 
P}(i),5_i}\gamma^T_{\Omega{\bf 
P}(i),5_i}) + 32^2 \end{equation}
(c) for the $D1$-brane,
\begin{equation} Tr(\gamma_{0,1})^2 - 64Tr(\gamma^{-1}_{\Omega{\bf 
P}(i),1}\gamma^T_{\Omega{\bf P}(i),1}) + 
32^2 \end{equation}

Now using that the number of branes of a particular type is given by the trace 
of $\gamma_{1,p}$ we can solve 
these three equations to find that the number of each type of brane is always 32 
provided they exist in the 
relevant model to start with. This is a general result and is consistent with 
what we know about $D$--branes \cite{tasi}.

For the twisted sectors we have the results,\\
(a) from the Klein bottle.
\begin{equation}  
\left[ 2^{5-d}\sum_{{\bf P}(i)}\prime  \prod^d_i {2cos \pi z_i \over 2sin \pi 
z_i} \right]^2
\end{equation}
which for $\Omega_{k+ N/2}$ becomes 
\begin{equation} 
\left[2^{5-d}\sum_{{\bf P}(i)}\prime \prod^d_i {-sin\pi z_i \over sin \pi 
z_i}\right]^2
 \end{equation}
(b) from the M\"{o}bius strip
\begin{eqnarray}
-2^{6-d}\sum_{{\bf P}(i)}\prime \left\{ 
Tr(\gamma^{-1}_{\Omega,9}\gamma^T_{\Omega,9})\prod^d_i {1 \over 2sin\pi z_i} 
\right.
\nonumber \\
+Tr(\gamma^{-1}_{\Omega{\bf P}(i),5_i}\gamma^T_{\Omega{\bf P}(i),5_i})\prod^d_i 
(2cos\pi z_i)\nonumber  \\
\left. + Tr(\gamma^{-1}_{\Omega{\bf P}(i),1}\gamma^T_{\Omega{\bf 
P}(i),1})\prod^d_i 
(2cos\pi z_i)\right \} 
\end{eqnarray} 
(c) for the cylinder we must sum over all possible sectors. This can cause 
problems with the NN trace 
contributions but the expression can be written, after factorising into perfect 
squares, as:
\begin{eqnarray}
\sum_{{\bf P}(i)} \prime \left\{
\left(Tr(\gamma_{{\bf P}(i),9}) \prod^d_{i=1}(2sin\pi z_i)^{-1} -Tr(\gamma_{{\bf 
P}(i),5_i}){\prod^d_{i=1}(2sin\pi z_i) \over \tilde{\prod}(2sin\pi z_i)} 
\right)^2 \right.\nonumber\\
+\left(Tr(\gamma_{{\bf P}(i),9}) \prod^d_{i=1} (2sin\pi z_i)^{-1} - 
Tr(\gamma_{{\bf P}(i),1})\prod^d_{i=1}(2sin\pi z_i)\right)^2\nonumber\\
+\left(Tr(\gamma_{{\bf P}(i),5_i}) {\prod^d_{i=1}(2sin\pi z_i) \over 
\tilde{\prod}(2sin\pi z_i)} - 
Tr(\gamma_{{\bf P}(i),1})\prod^d_{i=1}(2sin\pi z_i)\right)^2\nonumber\\
+\left. \left(Tr(\gamma_{{\bf P}(i),5_i}) {\prod^d_{i=1}(2sin\pi z_i) \over 
\tilde{\prod}(2sin\pi z_i)} - Tr(\gamma_{{\bf P}(i),5_j})^2 
{\prod^d_{i=1}(2sin\pi z_i) \over \tilde{\prod}(2sin\pi z_i)}\right)^2 \right\}
\end{eqnarray}
where the tilde means that the contribution is from the coordinates were the 
branes do not overlap in the 
compact dimensions (ie ND coordinates) and the prime indicates we do not sum 
over the order two element. Note 
also that there are sums over the fixed points in the case of the $5_1$ and 1 
branes. 

Before we start solving for the individual sectors and extracting the 
divergences from these tadpoles let us 
note some general properties of $\gamma_{\Omega{\bf P}(i)}$ that hold in all 
cases. First we note that the 
Abelian group elements $\alpha^k_N$ and $\alpha^{k+ N/2}_N$ both square to the 
same element $\alpha^{2k}_N$. 
Also in solving the contributions from the untwisted sector we made the choice 
$\gamma_{N/2}=-1$. Finally we 
note that the values of $\Omega^2$ in the various sectors are, following 
\cite{gp}, 1 for the 99, 91  and 11 
sectors, $-1$ for the $5_i5_i$, $5_i9$ and $5_i1$ sectors. Using this we can 
define a set of relations 
allowing us to simplify the above equations.
\begin{eqnarray}
Tr(\gamma^T_{\Omega k}\gamma^{-1}_{\Omega k}) & = & Tr(\gamma_{\Omega 2k}) \\
Tr(\gamma^T_{\Omega k+N/2}\gamma^{-1}_{\Omega k+N/2}) & = & -Tr(\gamma_{\Omega 
2k})
\end{eqnarray}
when $\Omega^2=1$; and 
\begin{eqnarray}
Tr(\gamma^T_{\Omega k}\gamma^{-1}_{\Omega k}) & = &-Tr(\gamma_{\Omega 2k}) \\
Tr(\gamma^T_{\Omega k+N/2}\gamma^{-1}_{\Omega k+N/2}) & = &Tr(\gamma_{\Omega 
2k})
\end{eqnarray}
when $\Omega^2=-1$ for the various brane combinations.

Going back to purely Abelian case and looking ahead somewhat, we know that 
depending on the particular 
elements of the point group, the cylinder equation above is what we get for some 
constraints while for others 
we have to take in to account the M\"{o}bius strip and Klein bottle 
contributions. Assuming that we have to 
do the latter notice that if we take this charge squared expression to be of the 
following form and do a 
little algebra:
\begin{equation} 
(a - b - c)^2 = (a-b)^2 + c^2 - 2c(a-b) 
\end{equation}
we can immediately take $(a-b)^2$ as the contribution from the cylinder, $c^2$ 
as the contribution from the Klein bottle and the M\"{o}bius strip must be twice 
the square root of the Klein bottle times the square root of the cylinder 
contribution $(a-b)$. An examination of the amplitudes shows that this is indeed 
holds generally. We can do this due to the Gimon--Polchinski conditions which 
mean that the Klein bottle and M\"{o}bius strip will not contribute in every 
twisted sector so that the cylinder expression must form a perfect square on its 
own. So whereas before the Gimon--Polchinski conditions seemed to imply more 
work they now give the remarkable result of simplifying the expressions for the 
purely Abelian case. 

Apply this we get the set of charge contraints for arbitary $d$, 
\begin{eqnarray} 
\sum_{{\bf P}(i)} \prime\left(Tr(\gamma_{{\bf P}(i),9}) \prod^d_{i=1}(2sin\pi 
z_i)^{-1} -Tr(\gamma_{{\bf 
P}(i),5_i}){\prod^d_{i=1}(2sin\pi z_i) \over \tilde{\prod}(2sin\pi z_i)}- K 
\right)^2 \\
\sum_{{\bf P}(i)} \prime \left(Tr(\gamma_{{\bf P}(i),9}) \prod^d_{i=1} (2sin\pi 
z_i)^{-1} - Tr(\gamma_{{\bf 
P}(i),1})\prod^d_{i=1}(2sin\pi z_i) - K \right)^2 
\end{eqnarray}
\begin{eqnarray}
\sum_{{\bf P}(i)} \prime \left(Tr(\gamma_{{\bf P}(i),5_i}) 
{\prod^d_{i=1}(2sin\pi z_i) \over 
\tilde{\prod}(2sin\pi z_i)} - Tr(\gamma_{{\bf P}(i),1})\prod^d_{i=1}(2sin\pi 
z_i) - K \right)^2 \\
\sum_{{\bf P}(i)} \prime \left(Tr(\gamma_{{\bf P}(i),5_i}) 
{\prod^d_{i=1}(2sin\pi z_i) \over 
\tilde{\prod}(2sin\pi z_i)} - Tr(\gamma_{{\bf P}(i),5_j})^2 
{\prod^d_{i=1}(2sin\pi z_i) \over 
\tilde{\prod}(2sin\pi z_i)} - K \right) ^2
\end{eqnarray}
where we have defined $K$ as
\begin{equation}
2^{5-d}\left\{ \prod^d_{i=1}{2cos\pi z_i \over 2sin\pi z_i} + 
\prod^d_{i=1}{-2sin\pi z_i \over 2sin\pi z_i} \right\}
\end{equation}
corresponding to the $\Omega_k$ and $\Omega_{k+N/2}$ Klein bottle amplitudes 
respectively. Each expression is individually set to zero. What we are looking 
for is that the total $D$-brane plus orientifold charge squared in each twisted 
sector is zero. Remarkably this can be solved as it stands if we make the 
subsitutions
\begin{eqnarray} 
2cos\pi z_i & = & e^{\pi i z_i} + e^{-\pi i z_i} \nonumber \\
2isin\pi z_i & = & e^{\pi iz_i} + e^{-\pi i (z_i + 1)} 
\end{eqnarray}
First look at the $D9$--branes, pulling out the common factor of $(2sin\pi 
z_i)^{-1}$. As a result the factor in front of the $D5$--branes and $D1$--branes 
is related to the Lefschetz fixed point formula for the appropriate element of 
${\bf P}(i)$. In the cases we will be interested in this is infact an integer. 
We now let 
\begin{equation}
Tr(\gamma_{{\bf P}(i),9}) = Tr(\gamma_{{\bf P}(i),1})
\end{equation}
which implies that the trace of $\gamma_{{\bf P}(i),9}$ is equal to the 
contribution from the Klein bottle up to an integer. It is worth looking at the 
Klein bottle contribution a little closer. For clarity let us change the 
notation so that we have instead $\pi z_1=a, \pi z_2=b, \pi z_3=c, \pi z_4=d$, 
and assume we have a hypothetical $\pi z_5=f$. Calculating $(2cos\pi z_i)^d$ for 
$d=2, 3, 4$ we get the corresponding pattern:
\begin{eqnarray}
e^{i(\pm a\pm b)} & =&  4\,\,terms \nonumber \\
e^{i(\pm a\pm b \pm c)} & =&  8\,\, terms \nonumber \\
e^{i(\pm a\pm b \pm c \pm d)} & =&  16\,\, terms \nonumber \\
e^{i(\pm a\pm b \pm c \pm d \pm f)} & =&  32\,\, terms
\end{eqnarray}
Similiarly for $(2sin\pi z_i)^d$.

Each of these terms can be interpreted as the diagonal components of the 
$\gamma$ matrix associated to the $D9$--brane 
operating in the various sectors. They encode all the group properties. If we 
multiply the number of terms by 
the remaining coefficient of the Klein bottle contribution then we recover the 
number 32 in all cases as 
required to make up a trace of a matrix corresponding to 32 $D$--branes. If we 
did not have 32 terms then 
this would imply that we would have less than 32 $D$--branes which from the 
untwisted sector is not possible. 
This also essentially implies the uniqueness of the solution. 

Thus we can write the solution for the $D9$--brane in an arbitary sector as 
\begin{equation}
\gamma_{1,9} = diag\{ e^{i\pi(\xi)} \} \otimes {\bf I}_{2^k}
\end{equation}
where $(\xi)$ is all possible combinations of $\pm z_i$ and then all possible 
combinations of $\pm (z_i + 1)$ 
when they apply, ie even $N$, with $k$ being the appropriate value need to make 
up the number of terms to 32. 
What is more is that with this notation we do not need to worry about how the 
$\gamma$'s will look in a cross 
model when there are two or more point groups of the same type, as they will 
have different values for $a, b, 
c, d$ which give rise to the same terms but in a different order. We simply are 
required to choose the initial 
order for the $\pm$ signs. Notice that setting any one of $a, b, c, d$ to zero 
is equivalent to switching off 
the compactification in the assosiated dimensions, or equivalently moving into 
the untwisted sector for that 
dimension. 

As simple as making this choice appears it actually has quite far reaching 
consequences. The expansion shows that each value on the diagonal of the 
$\gamma$ and hence every $D$--brane feels the action of each point group in the 
model. The value of the diagonal terms will also determine whether a particular 
$D$--brane is dynamical or not, and also its relation to the other $D$--branes. 
Once set we cannot arbitarily change the order as that would amount to changing 
the action of the point group on the coordinates. This is the heart of the 
control the orientifold planes have on the $D$-branes. The charge cancellation 
equations also give us a way of solving the $\gamma$ for the 1 and $5_i$ 
sectors by using the solution for $\gamma_9$ but these need to be considered for 
the individual models. 

The purpose of including the hypothetical $z_5$ which would be the contribution 
if we compactified the $x^0, 
x^1$ coordinates reserved for the lightcone quantization is to show that to 
compactify further is not 
possible. This is as the number of terms would be greater than 32 so the 
equations are unsolvable, besides violating 
what we know from the completely untwisted sector. What this tell us is that the 
scheme sees the total number 
of dimensions and hence the superconformal anomaly which fixes it as would be 
expected, but that we are 
unable to use this to place any constraints on how the compactification works. 

Before moving onto specific examples, we should note something about the 
solution for $\gamma$ when $N$ is 
even. We can consider the $sin\pi /N$ term as infact being $cos\pi (N-1)/N$. In 
fact we can use this as an 
alternative method for constructing the same solutions building on the fact that 
the trace of $\gamma_{1,9}$ 
is zero while $\gamma_{2,9}$ is not. However, more importantly, that we have to 
include both the $\Omega_k$ 
and $\Omega_{k+N/2}$ gives rise to the fact that the solution will always 
factorise to something with a $Z_2$ 
structure.

Since the cases for $d \leq 3$ have already been dealt with in the literature we 
complete the tale by 
constructing the $\gamma$'s for $d=4$, that is compactification down to two 
dimensions. The work on the cases 
on containing $N_i=2, \,i=1, 2, 3$ has been done in \cite{fg} and as it's 
solutions are somewhat different in 
structure than for $N_i \neq 2$ we will not cover them here.

\subsection{The $T^8/Z_N$ Orientifold}
Here we have the action of the point group acting on all the coordinates at once 
so we can write $z_i = z$ 
for all $i$, with $z=n/N$. In this case we have no $D5$-branes, only $D1$-branes 
and $D9$-branes. We need to 
take care as to whether $N$ is odd or even and in the case of $N$ even whether 
$n$ is odd or even. Putting 
all this into our general solution we get the following results for even $N$
\begin{equation}
Tr(\gamma_{2k-1,9}) - (2sin{(2k-1)\pi \over N})^8 Tr(\gamma_{2n-1,1})  =0
\end{equation}
for $n=2k-1$; and for $n=2k$
\begin{equation} 
Tr(\gamma_{2k,9})-(2sin{2k\pi \over N})^8 Tr(\gamma_{2k,1})=2.\left(2cos({2k\pi 
\over N})\right)^4+\left( 2sin({2k\pi \over N})\right)^4 
\end{equation}
While for odd $N$
\begin{eqnarray}
 Tr(\gamma_{n,9}) - (2sin{n\pi \over N})^8 Tr(\gamma_{n,1}) - 2.\left(2cos({n\pi 
\over N})\right)^4 =0
\end{eqnarray}
These can now be solved to find $\gamma$.
First solve for the case when $N$ is odd. Since we have only one constraint per 
sector we set
\begin{equation}
Tr(\gamma_{n,9}) = Tr(\gamma_{n,1})
\end{equation}
which gives us
\begin{equation}
\gamma_{n,9} = diag\{ e^{4\pi iz}, e^{-4\pi iz}, e^{2\pi iz}(4), e^{-2\pi 
iz}(4), 1(6)\} \otimes {\bf I}_{2}
\end{equation}
 The number in parenthesis is the multiplicity of that particular value, we are 
not going to overly concern ourselves with the order as we are not going to 
evaluate the Chan--Paton factors or Wilson lines specifically. For the case 
$N=3$, we get:
\begin{eqnarray}
\gamma_{1,9} = diag\{ e^{\pi i/3}, e^{-\pi i/3},e^{2\pi i/3}(4), e^{-2\pi 
i/3}(4), 1(6) \} \otimes {\bf I}_{2}
\end{eqnarray}

When $N$ is even we have the solutions:
\begin{equation}
Tr(\gamma_{2k-1,9}) = Tr(\gamma_{2k-1,1}) = 0 
\end{equation}
which tells us that we can set $\gamma_{n,9}=e^{i\pi m/N}\gamma_{n,1}$ where $m$ 
is an odd integer, while the solution for $n=2k$ gives the result for  the 
diagonal of $\gamma_{1,9}$ 
\begin{eqnarray}
\{ e^{2\pi i/N}(2), e^{-2\pi i/N}(2), e^{\pi i/N}(4), e^{-\pi i/N}(4), -e^{\pi 
i/N}(4), -e^{-\pi i/N}(4), 1(12) \}
\end{eqnarray}
which for $N=2M$ has the factorisation $Z_N=Z_M \otimes Z_2$ as expected for 
$Z_4$ and $Z_6$ .

\subsection{The $T^8/(Z_N \times Z_M)$ Orientifold}
We define the action of the point group in this case as being 
\begin{equation}
\begin{array}{c c c c}
z:&\left(x^{2,3},x^{4,5},x^{6,7},x^{8,9}\right)&\to&\left(x^{2,3},x^{4,5},e^{-2\
pi iz}x^{6,7},e^{2\pi iz}x^{8,9}\right) \\
y:&\left(x^{2,3},x^{4,5},x^{6,7},x^{8,9}\right)&\to&\left(e^{-2\pi 
iy}x^{2,3},e^{2\pi iy}x^{4,5},x^{6,7},x^{8,9}\right)
\end{array}
\end{equation}
where $z=n/N$ and $y=m/M$ labels the action of $Z_N$ and $Z_M$. respectively.

We can see that this divides the compact space up into $T^4/Z_N \otimes T^4/Z_M$ 
but now due to the 
existence of two $D5$-branes, one for each invariant space left under the action 
of $Z_N$ or $Z_M$, it is not 
possible to treat them completely independently except perhaps for those point 
groups without an element of 
order two. However, we can use this fact to see that the tadpole divergences are 
evenly split between the two 
sectors with $z_1 = z_2 = z,\, z_3 = z_4= y$. We should also consider the cases 
when $n=0, m \neq 0$ and 
$m=0, n \neq 0$ seperately though following the prescription given earlier they 
can be obtained simply enough 
from the following equations; $n=0, m=0$ is just the untwisted sector. Define 
$5_1$ to be the $D5$--brane 
left invariant under the action of $z$ and $5_2$ to be the $D5$--brane left 
invariant under the action of 
$y$.

The tadpole contributions are for $n \neq 0, m \neq 0$:\\
(i) from the Klein bottle
\begin{equation} 
4\left[ {(2cos\pi z)^2(2cos\pi y)^2 \over (2sin\pi z)^2(2sin\pi y)^2} \right]^2  
\end{equation}
and a similiar contribution for the $\Omega_{(k+N/2)}$ diagram where as before 
the 
relevant topline $(2cos\pi z_i)$ goes to $(-2sin\pi z_i)$.\\
(ii) from the M\"{o}bius strip, $-4 \times$
\begin{eqnarray}
Tr(\gamma^{-1}_{\Omega,z,y,9}\gamma^T_{\Omega,z,y,9}) {1 \over (2sin\pi 
z)^2(2sin\pi y)^2} \nonumber \\
+ Tr(\gamma^{-1}_{\Omega z,y,5_1}\gamma^T_{\Omega z,y,5_1})(2cos\pi z)^2(2cos\pi 
y)^2  \nonumber \\
+ Tr(\gamma^{-1}_{\Omega z,y,5_2}\gamma^T_{\Omega z,y,5_2})(2cos\pi z)^2(2cos\pi 
y)^2  \nonumber \\
+ Tr(\gamma^{-1}_{\Omega z,y,1}\gamma^T_{\Omega z,y,1})(2cos\pi z)^2(2cos\pi 
y)^2 
\end{eqnarray}
(iii) from the cylinder
\begin{eqnarray}
\sum_{n,m}^{N-1 \atop M-1}\prime \sum_{(G)} \left\{  
\left( Tr(\gamma_{z,y,9})(2sin\pi z)^{-2}(2sin\pi y)^{-2} - 
Tr(\gamma^{(G)}_{z,y,1}) \right)^2 \right. \nonumber\\ 
+\left( Tr(\gamma_{z,y,9})(2sin\pi z)^{-2}(2sin\pi y)^{-2} - 
Tr(\gamma^{(G)}_{z,y,5_1})(2sin\pi z)^{2} \right)^2 \nonumber \\
+\left( Tr(\gamma_{z,y,9})(2sin\pi z)^{-2}(2sin\pi y)^{-2} - 
Tr(\gamma^{(G)}_{z,y,5_2})(2sin\pi y)^{2}\right)^2 \nonumber\\
+\left( Tr(\gamma^{(G)}_{z,y,5_1})(2sin\pi z)^{2} - Tr(\gamma^{(G)}_{z,y,1}) 
\right)^2 \nonumber\\   
+\left( Tr(\gamma^{(G)}_{z,y,5_2})(2sin\pi y)^{2} - Tr(\gamma^{(G)}_{z,y,1}) 
\right)^2 \nonumber\\
\left. +\left( Tr(\gamma^{(G)}_{z,y,5_1})(2sin\pi z)^{2} - 
Tr(\gamma^{(G)}_{z,y,5_2})(2sin\pi y)^{2} \right)^2 \right\}   
\end{eqnarray}
where the $(G)$ is the sum over fixed points. This can be solved as it stands 
using the method outlined 
already, or we can switch a $z$ or $y$ off and see that we simply have the 
solutions for $T^4/Z_N$ which are 
already presented in \cite{gj} which can be multiplied together as desired to 
obtain the relative sectors. 
All we have to be aware of is the phases between the $5_i$ and 1 branes which 
will be the same as given in 
\cite{fg}.

\subsection{The $T^8/(Z_N \times Z_M \times Z_P)$ Orientifold}
We define the action of the point group as being
\begin{equation}
\begin{array}{c c c c}
z:&\left(x^{2,3},x^{4,5},x^{6,7},x^{8,9}\right)&\to&
\left(x^{2,3},x^{4,5},e^{2\pi i z}x^{6,7},e^{-2\pi i z}x^{8,9}\right)\\
y:&\left(x^{2,3},x^{4,5},x^{6,7},x^{8,9}\right)&\to&
\left(e^{2\pi i y}x^{2,3},e^{-2\pi i y}x^{4,5},x^{6,7},x^{8,9}\right)\\
x:&\left(x^{2,3},x^{4,5},x^{6,7},x^{8,9}\right)&\to&
\left(e^{-2\pi i x}x^{2,3},x^{4,5},e^{2\pi i x}x^{6,7},x^{8,9}\right)\\
\end{array}
\end{equation}
with $z=n/N$, $y=m/M$ and $x=p/P$ labels the action of $Z_N$, $Z_M$ and $Z_P$. 
respectively. This case will 
have all brane types, but requires that 
\begin{equation} {N \over M} \in {\bf Z},\,\,\,{M \over P} \in {\bf Z} 
\end{equation}
 so restricting the number of possible combinations of point groups somewhat, 
eg. we can have 2,2,2 or 3,3,6 
but not 2,3,6 etc. for $N,M,P$ respectively. As in the previous case most of the 
work has already been done 
for us in \cite{z} with the phases again from \cite{fg}. All we have to do is 
switch off all point groups 
except for one, which effectively leaves us with a $T^4/Z_N$ to solve again. 
What makes this one different is 
the ordering of terms in the solutions for $\gamma$ but this is done by 
comparison with the general solution 
presented earlier. As will be discussed in the next section models containing 
one or more $Z_4$'s will be 
inconsistent.

Other types of $d=4$ models can be constructed by playing around with the basic 
constructions presented here, 
such as a $T^8/Z^4_{N_i}$ where 
\begin{equation}
\begin{array}{c c c c}
z:&\left(x^{2,3},x^{4,5},x^{6,7},x^{8,9}\right)&\to&
\left(x^{2,3},x^{4,5},e^{2\pi i z}x^{6,7},e^{-2\pi i z}x^{8,9}\right)\\
y:&\left(x^{2,3},x^{4,5},x^{6,7},x^{8,9}\right)&\to&
\left(e^{2\pi i y}x^{2,3},e^{-2\pi i y}x^{4,5},x^{6,7},x^{8,9}\right)\\
x:&\left(x^{2,3},x^{4,5},x^{6,7},x^{8,9}\right)&\to&
\left(e^{-2\pi i x}x^{2,3},x^{4,5},e^{2\pi i x}x^{6,7},x^{8,9}\right)\\
w:&\left(x^{2,3},x^{4,5},x^{6,7},x^{8,9}\right)&\to&
\left(x^{2,3},x^{4,5},e^{-2\pi i w}x^{6,7},e^{-2\pi i w}x^{8,9}\right)
\end{array}
\end{equation}
which works for $Z_N$, $Z_M$, $Z_P$ and $Z_Q$ and 
\begin{equation} 
{N \over M} \in {\bf Z},\,\,\,{M \over P} \in {\bf Z},\,\,\,{P \over Q} \in {\bf 
Z}
\end{equation}
and so on. However, the solutions will be simply those already known with 
reordering of the terms as 
explained earlier.

\section{Zwart Inconsistencies}
In \cite{z} Zwart noticed that the $T^6/(Z_2\times Z_4)$ and $T^6/(Z_4 \times 
Z_4)$ orientifold models are 
inconsistent as the relevant $\gamma$ matrices do not satisfy the appropriate 
group properties required. We 
are now in a position to give an explaination of why this anomalous result 
occurs and how it might occur in 
other models. It can also be seen that part of the problem lies in the $Z_4$ 
orbifold model. However, the 
pure $T^6/Z_4$ model is consistent so another part of the problem lies in the 
fact that it is a cross model. 
From Berkooz and Leigh \cite{bl}, we know that the representation for $Z_2 
\times Z_2$ is unique.  

The key to this lies one of the results in the previous section, which was that 
we can decompose the $\gamma$ matrices corresponding to $Z_N$ with $N = 2M$ in 
to the representations for $Z_M \otimes Z_2$. Going back to the set of 
automorphisms for the two dimensional (complex) lattice from which we get the 
available Abelian point groups we notice that the maximum point group for the 
$SU(2) \times SU(2)$ and $SU(3)$ lattices are both even. We can immediately see 
that the $Z_2$ in our decompostion corresponds to the general result for Lie 
algebra roots that every root has a negative value that is also a root and the 
$Z_2$ is simply the transformation between them:
\begin{equation}
Z_2: \alpha \rightarrow -\alpha
\end{equation}
This is a feature of all even point groups and thus we can take it as given. The 
problem must therefore lie in the remaining part of the point group. 

Let us stay with the $T^6$ models of Zwart. We see that models with $Z_6$ work 
fine. $Z_6$ splits up into 
$Z_3 \times Z_2$. Divide the root diagram into the six component triangles, the 
Weyl chambers, delineated by 
the various root vectors. The lattice we start with are the first two triangles 
between the two simple roots. 
The action of $Z_3$ is to rotate the root vectors to the third and fourth 
triangles and then the fifth and 
sixth triangles in turn. In each case we have the same lattice structure so thus 
the same physics, but the 
lattice has a different orientation relative to the original lattice. This 
difference in orientation means we 
can not identify the lattices under the lattice property.  Lattice rotations 
under the $Z_2$ rotation of the 
simple roots is trivially identified with the original lattice.

Doing the same for $Z_4$ which in terms of the $\gamma$'s has a $Z_2 \times Z_2$ 
decomposition. As before one $Z_2$ corresponds to the interchange of the 
positive roots with their negative counterparts and the new lattice formed is 
trivially identified with the original one. The second $Z_2$, call it 
$Z_2^\prime$, corresponds to a reflection in the axes, but as this is now the 
$SU(2) \times SU(2)$ root system these are orthogonal so they divide the space 
up into four identical squares that under the lattice identification can all be 
mapped onto each other. Also the model must contain the $Z_2 \times Z_2$ 
solution in it. Acting with the $\gamma$ matrix corresponding to $Z_2^\prime$ on 
the 
Berkooz--Leigh solution swaps the components of the Berkooz--Leigh solution 
around. However, we must be able 
to identify the $\gamma$ matrices of the Berkooz--Leigh solution with the new 
matrices due to the fact that 
they are identified under the lattice. An examination of the solutions quickly 
shows that it is not possible 
with out making the identification $\epsilon = -\epsilon$ for at least one of 
the sectors. 

Thus the origin of the inconsistancy can be taken back directly to the 
uniqueness of the Berkooz--Leigh 
solution for the $Z_2 \times Z_2$ model. That this solution is unique stems from 
the fact that the $Z_2$'s 
involved correspond to the interchanging of the positive and negative simple 
roots in each $T^2$ the complex 
coordinate system divides up the compact space into. This with the fact it is a 
cross model is what makes the solution 
unique so the same analysis is not directly applicable to the pure $Z_2$ models.

We must also note that this does not appear to be a feature of simple orbifold 
models but arise out of the 
fact we are taking T--dual models and enhancing the point group to contain a 
worldsheet parity, which we can 
assosiate the $Z_2$ with \cite{dp1} \cite {dp2}.

Any orientifold model whose component lattices shows similiar behavour to the 
$Z_4$ cross models will suffer 
the same fate of having a inconsistent representation of the action for the 
point group on the Chan Paton 
factors. Thus we can also expect similar behaviour to be observed in the 
solutions of orientifold 
compactification on $T^8$.
 
\section{Discussion}
In this paper we have examined the general orientifold model on the $SU(N)$ 
subgroups of the $SO(N)$ space 
groups in compactifying the type IIB superstring down to 6, 4 and 2 dimensions. 
What we have discovered is 
that in all these models there must always be 32 $D$--branes in them. We can 
also see that the dynamics of 
the $D$--branes is much more strongly controlled by the Klein bottle 
contributions to the amplitudes, and 
thus by the orientifold planes, than was suspected. Rather than being present to 
ensure proper charge 
cancellation, they impose stringent conditions on the form of the $\gamma$'s 
representing how the point group 
act on the $D$--branes.

It is possible to generalise these models even further. So far all work in the 
area has been limited to the 
$SU$ group lattices and Abelian point groups. This has been because they have 
supersymmetry preserving 
properties and can be related to special points in Calabi--Yau moduli spaces. 
However, it is possible to 
consider more general models that have features such as discrete torsion, 
asymmetry or non--Abelian point 
groups. While the former two can be considered for the models examined in this 
paper the non--Abelian point 
groups require a fresh approach. The $\vartheta$ notation used in this paper 
simplified the amplitudes 
considerable but they are only suitable for a fermion on a complex manifold. For 
non--Abelian group we would 
want to allow for real manifold so have to retreat to a more basic form 
utilising the $f$ functions modified 
to take a $exp(2 \pi iz)$ term \cite{donal}. 

Another approach that needs to be explored more fully is the use of boundary 
operators, which have been used to describe $D$--branes. Analogous to them are 
crosscaps \cite{cp, dunbar88} which in the T--dual model may give a description 
of 
the orientifold planes. A test for them would be that they should be able to 
reproduce all the amplitudes presented here. This would be useful as they are 
easier to construct for non--standard point groups.

After this paper was finished refs \cite{ks1, ks2} were brought to our 
attention. In particular the treatment of the tadpoles and the construction of 
the $\gamma$'s \cite{ks1} overlaps somewhat with ours, and corroborates many of 
the results which we have obtained. Indeed they have considered Abelian groups 
not explicitly presented in this paper, though it is obvious they can be readily 
accounted for by our results.
\newline
{\bf Acknowlegments}\\
We would like to say a particular thank you to D. Dunbar, S. F\"{o}rste and T. 
Hollowood; and also acknowledge useful communications with D. Ghoshal, C. 
Johnson and G. 
Zwart. This work was supported by PPARC.

\end{document}